%% file: draft_BAM245_final.tex
\documentclass[amsmath,amssymb,twocolumn,prd,floatfix,showpacs, nofootinbib]{revtex4-1}
\usepackage{tabularx}  
\usepackage{bm, color} 
\usepackage{overpic,subfigure} 
\usepackage{multirow}
\usepackage{array}
\usepackage{dcolumn} 
\usepackage[symbol]{footmisc}

\begin{document}
\newcommand{\bes}{\mbox{BES\uppercase\expandafter{\romannumeral3}} }
\newcommand{\mbc}{$M_\mathrm{BC}$ }
\newcommand{\dplus}{$D^+$ }
\newcommand{\dzero}{$D^0$ }
\newcommand{\dele}{$\Delta E$ }
\newcommand{\kshort}{$K_S^0$ }
\newcommand{\etap}{$\eta^\prime$ }
\newcommand{\pizpip}{$D^+ \rightarrow \pi^0 \pi^+$ }
\newcommand{\pizkp}{$D^+ \rightarrow \pi^0 K^+$ }
\newcommand{\etapip}{$D^+ \rightarrow \eta \pi^+$ }
\newcommand{\etakp}{$D^+ \rightarrow \eta K^+$ }
\newcommand{\etappip}{$D^+ \rightarrow \eta^\prime \pi^+$ }
\newcommand{\etapkp}{$D^+ \rightarrow \eta^\prime K^+$ }
\newcommand{\kspip}{$D^+ \rightarrow K_S^0 \pi^+$ }
\newcommand{\kskp}{$D^+ \rightarrow K_S^0 K^+$ }
\newcommand{\ksetap}{$D^0 \rightarrow K_S^0 \eta^\prime$ }

\title{\boldmath Measurements of absolute branching fractions for $D$ mesons decays into two pseudoscalar mesons}
\input{./author_final}

\begin{abstract}
Using a data sample of $e^+e^-$ collision data with an integrated luminosity of 2.93 fb$^{-1}$ taken at the center-of-mass energy $\sqrt s= 3.773$~GeV with the BESIII 
detector operating at the BEPCII storage rings, we measure the absolute branching fractions of the two-body hadronic decays $D^+\to \pi^+\pi^0$, $K^+ \pi^0$, $\pi^+ \eta$, $K^+\eta$, $\pi^+\eta^\prime$, $K^+\eta^\prime$, $K_S^0 \pi^+$, $K_S^0 K^+$, and $D^0\to \pi^+ \pi^-$,
$K^+ K^-$, $K^\mp \pi^\pm$, $K_S^0 \pi^0$, $K_S^0 \eta$, $K_S^0 \eta^\prime$.
Our results are consistent with previous measurements within uncertainties.
Among them, the branching fractions for $D^+\to\pi^+\pi^0$, $K^+\pi^0$, $\pi^+\eta$,  $\pi^+\eta^\prime$,
$K_S^0 \pi^+$, $K_S^0 K^+$ and $D^0 \to K_S^0 \pi^0$, $K_S^0 \eta$, $K_S^0 \eta^\prime$ are determined with improved
precision compared to the world average values.
\end{abstract}
\pacs{13.25.Ft}

\maketitle

\renewcommand{\thefootnote}{\fnsymbol{footnote}}

\section{Introduction}

The two-body hadronic decays $D\to P_1P_2$ (throughout the text, $D$ represents the $D^+$ and $D^0$ mesons and $P$ 
denotes one of the pseudoscalar mesons 
$\pi^\pm$, $K^\pm$, $K^0_S$, $\pi^0$, $\eta$ and $\eta^\prime$)
serve as an ideal testbed to improve the understanding of the 
weak and strong interactions
in decays of charmed mesons. 
These reactions proceed via external $W\text{-emission}$, internal $W\text{-emission}$
or $W\text{-exchange}$ processes.
Due to the relatively simple topology, the amplitude of $D\to P_1P_2$ decay 
can be theoretically derived as a sum of different diagrams
based on SU(3)-flavor symmetry~\cite{chau1987analysis}.
Comprehensive and improved experimental measurements
of the branching fractions
for these decays may help to validate the
theoretical calculations and
provide important and complementary data to
explore the effect of SU(3)-flavor symmetry breaking in hadronic decays
of the $D$ mesons~\cite{waikwok1993minimal, Grossman2013,yu2011nonleptonic,li2012branching}.

Historically, experimental studies of
singly or doubly-Cabibbo-suppressed (DCS)
decays of $D\to P_1P_2$ with branching fractions at the $10^{-4}$ level
were challenging due to limited statistics and high background.
In recent years, the $D\to P_1P_2$ decays have
been widely studied in various experiments~\cite{cdf2005measurement,babar2006measurement, cleo2010measurements, belle2011observation, pdg2016review}.
The \bes Collaboration has recently reported measurements of the branching
fractions for 
some $D^{+(0)}\to P_1 P_2$ decays~\cite{hajime2015pipi, ke:omega, yue:ks, jingzhen:ks}
by analyzing the data sample corresponding to an integrated luminosity of $2.93\ \mathrm{fb^{-1}}$~\cite{ablikim2013measurement}
taken at the center-of-mass energy $\sqrt s= 3.773\ \mathrm{GeV}$.
Single-tag or double-tag methods, in which one or two $D$ mesons are
fully reconstructed, have been used in previous works.
Analyzing the same data sample with the single-tag method,
we report in this paper the measurements of the absolute branching fractions of the two-body
hadronic decays $D^+\to \pi^+\pi^0$,
$K^+\pi^0$, $\pi^+ \eta$, $K^+\eta$, $\pi^+\eta^\prime$, $K^+\eta^\prime$,
$K_S^0 \pi^+$, $K_S^0 K^+$, and $D^0\to \pi^+ \pi^-$,
$K^+ K^-$, $K^\mp \pi^\pm$, $K_S^0 \pi^0$, $K_S^0 \eta$, $K_S^0 \eta^\prime$,
where $D^0 \to K^\mp \pi^\pm$ includes both the Cabbibo-favored decay of $D^0 \to K^-\pi^+$ 
and the DCS decay of $D^0 \to K^+\pi^-$.
Throughout this paper, charge conjugated modes are implied.

\section{\bes detector and Monte Carlo simulation}

The \bes
detector is a cylindrical detector with a solid-angle coverage of 93\% of $4\pi$
that operates at the BEPCII collider. It consists of several main components.
A 43-layer main drift chamber (MDC) surrounding the beam pipe performs precise determinations
of charged particle trajectories and provides a measurement of ionization energy loss ($dE/dx$)
that is used for charged particle identification (PID). An array of time-of-flight counters
(TOF) is located outside the MDC and provides further information
for PID. A CsI(Tl) electromagnetic calorimeter (EMC)
surrounds the TOF and is used to measure the energies of photons and electrons. 
A solenoidal superconducting magnet outside the EMC
provides a $1\ \mathrm{T}$ magnetic field in the central tracking region of the
detector. The iron flux return yoke of the magnet is instrumented with about
$1272\ \mathrm{m^2}$ resistive plate muon counters (MUC), arranged in nine
layers in the barrel and eight layers in the end-caps, that are used
to identify muons with momenta greater than $0.5\ \mathrm{GeV}/c$.
More details about the \bes detector are described in Ref.~\cite{ablikim2010design}.

A GEANT4-based \cite{agostinelli2003geant4} Monte Carlo (MC) simulation software
package, which includes the geometric description of the detector
and its response, is used to determine the detection efficiency and
to estimate the potential background. An inclusive MC sample, which
includes $D^0\bar D^0$, $D^+D^-$ and non-$D\bar D$ decays of the
$\psi(3770)$, Initial State Radiation (ISR) production of the
$\psi(3686)$ and $J/\psi$, $e^+e^-\to q\bar q$ ($q=u$, $d$, $s$) continuum
processes, Bhabha scattering events, di-muon events and 
di-tau events, is produced at $\sqrt s=3.773\ \mathrm{GeV}$. The $\psi(3770)$
production is simulated by the MC generator KKMC~\cite{jadach2001coherent}, in which
the effects of ISR~\cite{kuraev1985radiative} and Final
State Radiation (FSR)~\cite{campbell1993supersymmetric} are considered.
The known decay modes are generated using
\mbox{EvtGen} \cite{lange2001evtgen} with the branching fractions taken from the
Particle Data Group (PDG)~\cite{[][ and 2011 partial update for the 2012 edition.]pdg2010review}, and
unknown decay modes are generated using LundCharm~\cite{chen2000event}.

\section{Data analysis}

The $D$ meson candidates are selected from combinations of $\pi^\pm$,
$K^\pm$, $K^0_S$, $\pi^0$, $\eta$ and $\eta^\prime$,
where $K^0_S$, $\pi^0$, $\eta$ and $\eta^\prime$ are 
reconstructed through their prominent decays $K^0_S\to \pi^+\pi^-$,
$\pi^0 \rightarrow \gamma \gamma$, $\eta \rightarrow \gamma \gamma$
and $\eta^\prime\to\pi^+\pi^-\eta$,
respectively.

All charged tracks, except for those from a $K^0_{S}$ decay, 
are required to originate from the interaction region
defined as $V_{xy}<1\ \mathrm{cm}$ and $|V_{z}|<10\ \mathrm{cm}$,
where $V_{xy}$ and $|V_z|$ denote the distances of the closest approach
of the reconstructed track to the Interaction Point (IP) in the
$xy$ plane and in the $z$ direction (along the beam direction), respectively.
The polar angles of the charged tracks $\theta$ 
is required to satisfy $|\mathrm{cos}\theta| < 0.93$.
Charged tracks are identified using confidence levels for the kaon (pion)
hypothesis $CL_{K(\pi)}$, calculated with both $dE/dx$ and TOF information.
The kaon (pion) candidates are required to satisfy $CL_{K(\pi)}>CL_{\pi(K)}$
and $CL_{K(\pi)}>0$.

The $K^0_S$ candidates are formed from two oppositely charged tracks with
$|V_z|<20\ \mathrm{cm}$ and $|\mathrm{cos}\theta| < 0.93$.
The two charged tracks are assumed to be a $\pi^+\pi^-$ pair without PID
and are constrained to originate from a common decay vertex.
To suppress the $\pi^+\pi^-$ combinatorial background,
the reconstructed decay length of the $K^0_S$ candidate is required to be greater than
twice its uncertainty.
The $\pi^+ \pi^-$ invariant mass must be within the signal region,
defined as $\pm 0.012\ \mathrm{GeV}/c^{2}$
around the $K^0_S$ nominal mass~\cite{pdg2016review}.

The photon candidates are selected from isolated EMC clusters.
To suppress the electronics noise and beam background, the clusters
are required to start within $700\ \mathrm{ns}$ after the event start time and 
fall outside a cone angle of $10^\circ$ around the nearest extrapolated charged track.
The minimum energy of each EMC cluster is required to be larger than 25 MeV
in the barrel region ($|\cos\theta|<0.80$) or 50 MeV in the end-cap region
($0.86<|\cos\theta|<0.92$)~\cite{ablikim2010design}.
To select the $\pi^0$ and $\eta$ meson candidates, the $\gamma\gamma$ invariant mass is required to be within
$(0.115, 0.150)\ \mathrm{GeV}/c^2$ and $(0.515, 0.575)\ \mathrm{GeV}/c^2$, respectively.
The momentum resolution of $\pi^0$ and $\eta$ is further improved with a kinematic fit that constrains the $\gamma\gamma$ invariant mass 
to the $\pi^0$ or $\eta$ nominal mass~\cite{pdg2016review}.
For $\eta^\prime$ mesons, the $\pi^+\pi^-\eta$ invariant mass is required to be
within the signal region, which is $\pm 0.012\ \mathrm{GeV}/c^{2}$ around the nominal $\eta^\prime$ mass~\cite{pdg2016review}.

For $D^0$ decays to $\pi^+ \pi^-$, $K^+ K^-$ and $K^\mp \pi^\pm$, the backgrounds arising from
cosmic rays and Bhabha events are rejected with the same requirements as those used
in Ref.~\cite{Ablikim:2014gvw}, \textit{i.e.}, the two charged tracks must have a TOF
time difference less than $5\ \mathrm{ns}$ and must not be consistent with the requirement for a muon
pair or an electron-positron pair. Furthermore, at least one EMC cluster 
with an energy larger than $50\ \mathrm{MeV}$ or at least one additional charged track detected
in the MDC is required.

At the $\psi(3770)$ peak, the $D\bar{D}$ meson pairs are produced without additional particles, thus,
the energies of the $D$ mesons are equal to the beam energy $E_\mathrm{beam}$ in the center-of-mass frame of the $e^+e^-$ system.
Two variables reflecting energy and momentum conservation are used to identify the $D$ meson candidates.
They are the energy difference
\begin{equation}
    \Delta E \equiv \sum_i E_i - E_\mathrm{beam},
\end{equation}
and the beam-energy-constrained mass
\begin{equation}
 M_\mathrm{BC}\cdot c^2 \equiv \sqrt{E^2_\mathrm{beam} - (\sum_i \vec{p}_i \cdot c)^2}, 
\end{equation}
where $E_i$ and $\vec{p}_i$ are the energy and momentum of the decay
products of the $D$ candidates in the center-of-mass frame of the $e^+e^-$ system. For a given $D$ decay mode, if there is more 
than one candidate per charm per event, the one with the least $|\Delta E|$ is kept 
for further analysis. The combinatorial backgrounds are suppressed by mode dependent $\Delta E$ requirements,
which correspond to $\pm 3.0\sigma_{\Delta E}$ around the fitted $\Delta E$ peak, where $\sigma_{\Delta E}$ is the
resolution of the $\Delta E$ distribution.

\begin{figure*}[htbp]
\centering
\begin{overpic}[width=.92\textwidth]{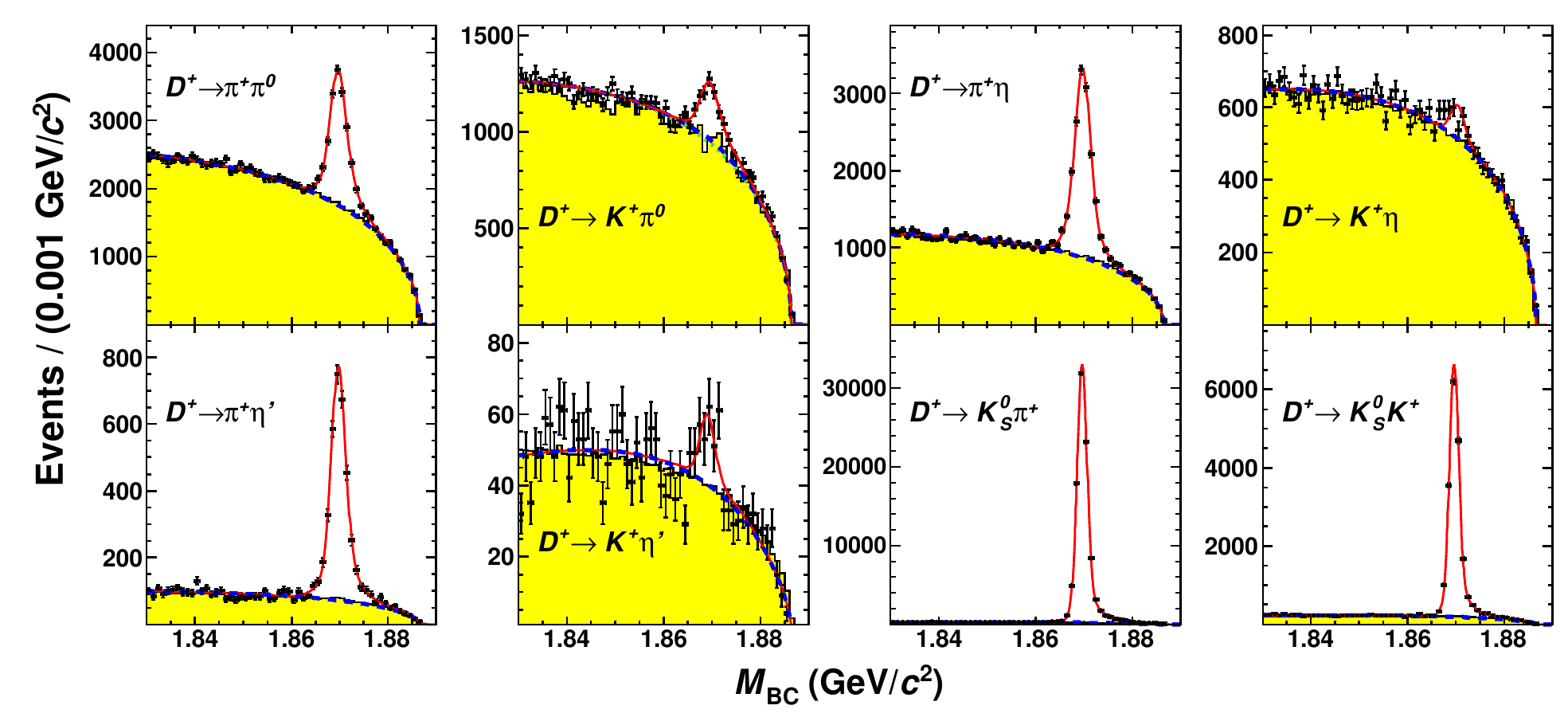}
\end{overpic}
\caption{
(Color online) Fits to the $M_\mathrm{BC}$ distributions of the single-tag $D^+$ candidate events.
The points with error bars are data,
the red curves are the overall fits,
the blue dashed curves are the fitted backgrounds and
the yellow shaded histograms are the MC-simulated combinatorial backgrounds.
For $D^+\to K^+\pi^0$, the blue dashed curve is the sum of the peaking background 
and the fitted ARGUS background.
}
\label{f_dp}
\end{figure*}

\begin{figure*}[htbp]
\centering
\begin{overpic}[width=.69\textwidth]{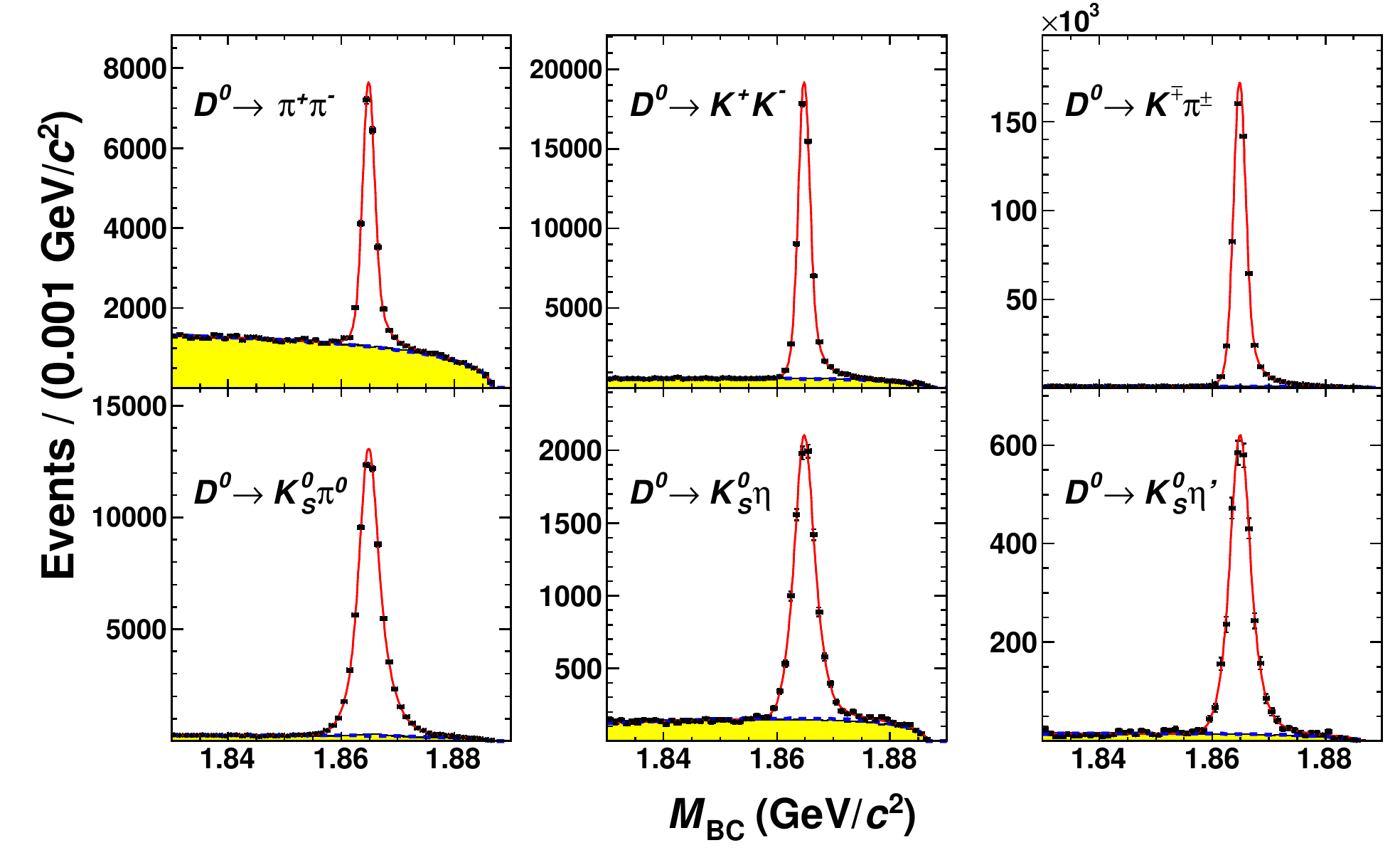}
\end{overpic}
\caption{
(Color online) Fits to the $M_\mathrm{BC}$ distributions of the single-tag $D^0$ candidate events.
The points with error bars are data,
the red solid curves are the overall fits,
the blue dashed curves are the fitted backgrounds and
the yellow shaded histograms are the MC-simulated combinatorial backgrounds.}
\label{f_d0}
\end{figure*}

Figures 1 and 2 show the $M_\mathrm{BC}$ distributions of 
the accepted single-tag $D^+$ and $D^0$ candidates, respectively.
The signal yields of $D$ mesons for the different processes are determined using 
unbinned maximum likelihood fits to the corresponding distributions,
where the signal probability density function (PDF) is modeled by the MC-simulated shape
convolved with a double Gaussian function that describes the resolution difference between data and MC simulation.
The combinatorial background is described with an ARGUS function~\cite{albrecht1989search} with
the endpoint fixed at $E_\text{beam}$.
For the DCS decay $D^+\to K^+\pi^0$, MC studies show that the sizeable peaking background
from $D^+\to K^0_S(\to\pi^0\pi^0)\pi^+$ can not be ignored. 
Thus, in the $M_\mathrm{BC}$ fit for this decay, the size and 
shape of the background $D^+\to K^0_S(\to\pi^0\pi^0)\pi^+$ are fixed based on MC simulation.

\begin{figure*}[htbp]
    \centering
    \subfigure[]{
        \begin{overpic}[width=.31\textwidth]{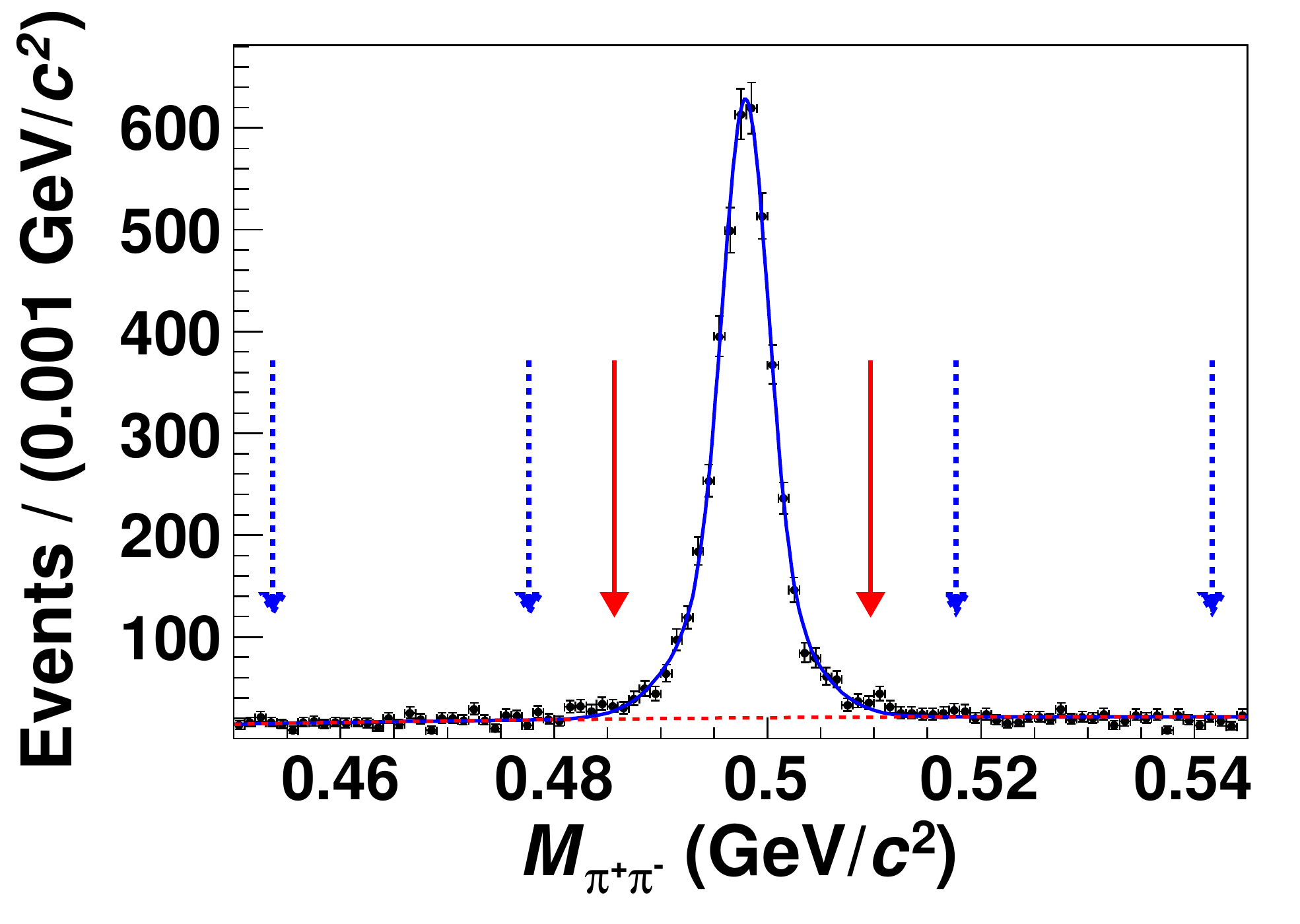}
        \end{overpic}
    }
    \subfigure[]{
        \begin{overpic}[width=.31\textwidth]{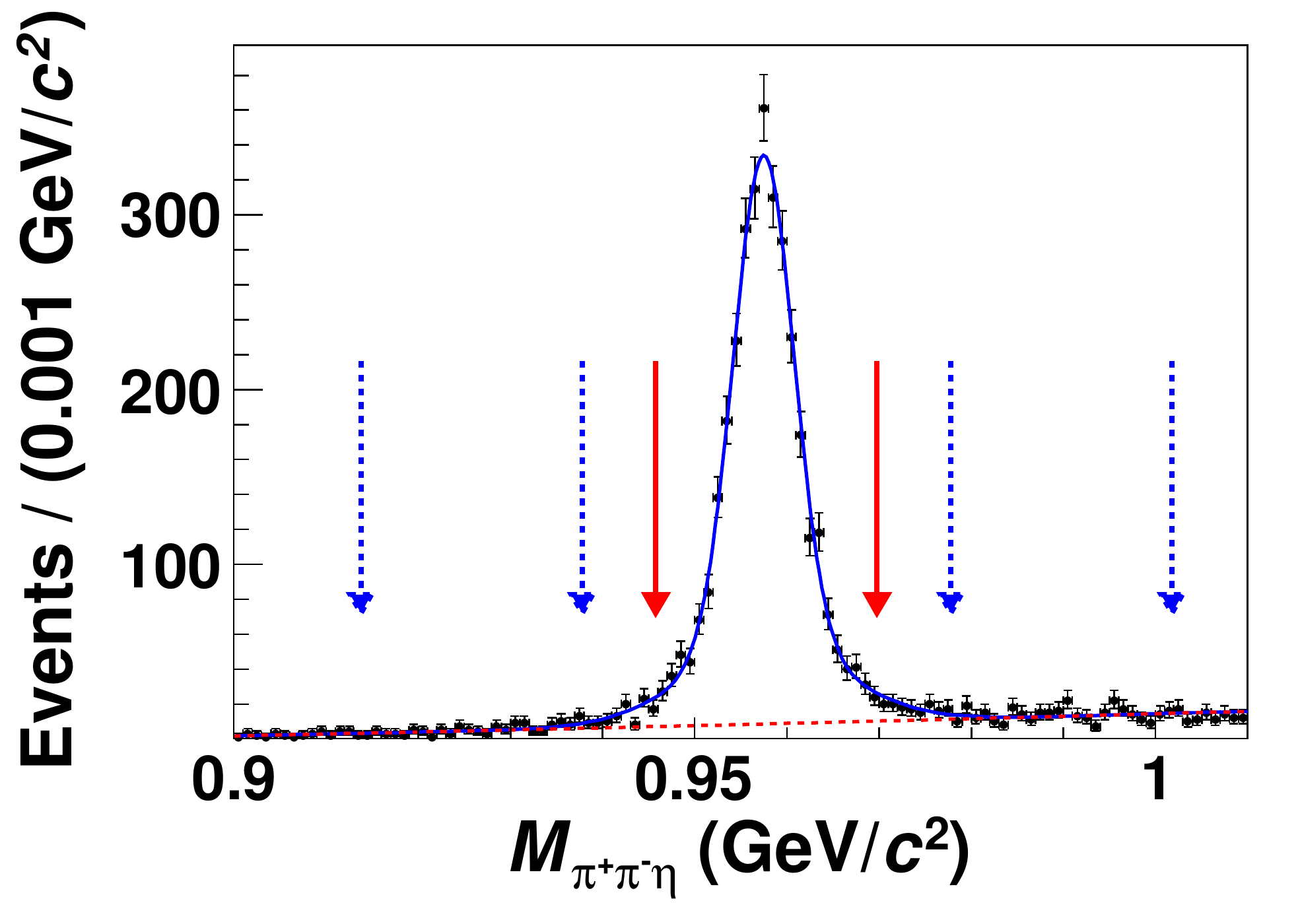}
        \end{overpic}
    }
    \subfigure[]{\begin{overpic}[width=.31\textwidth]{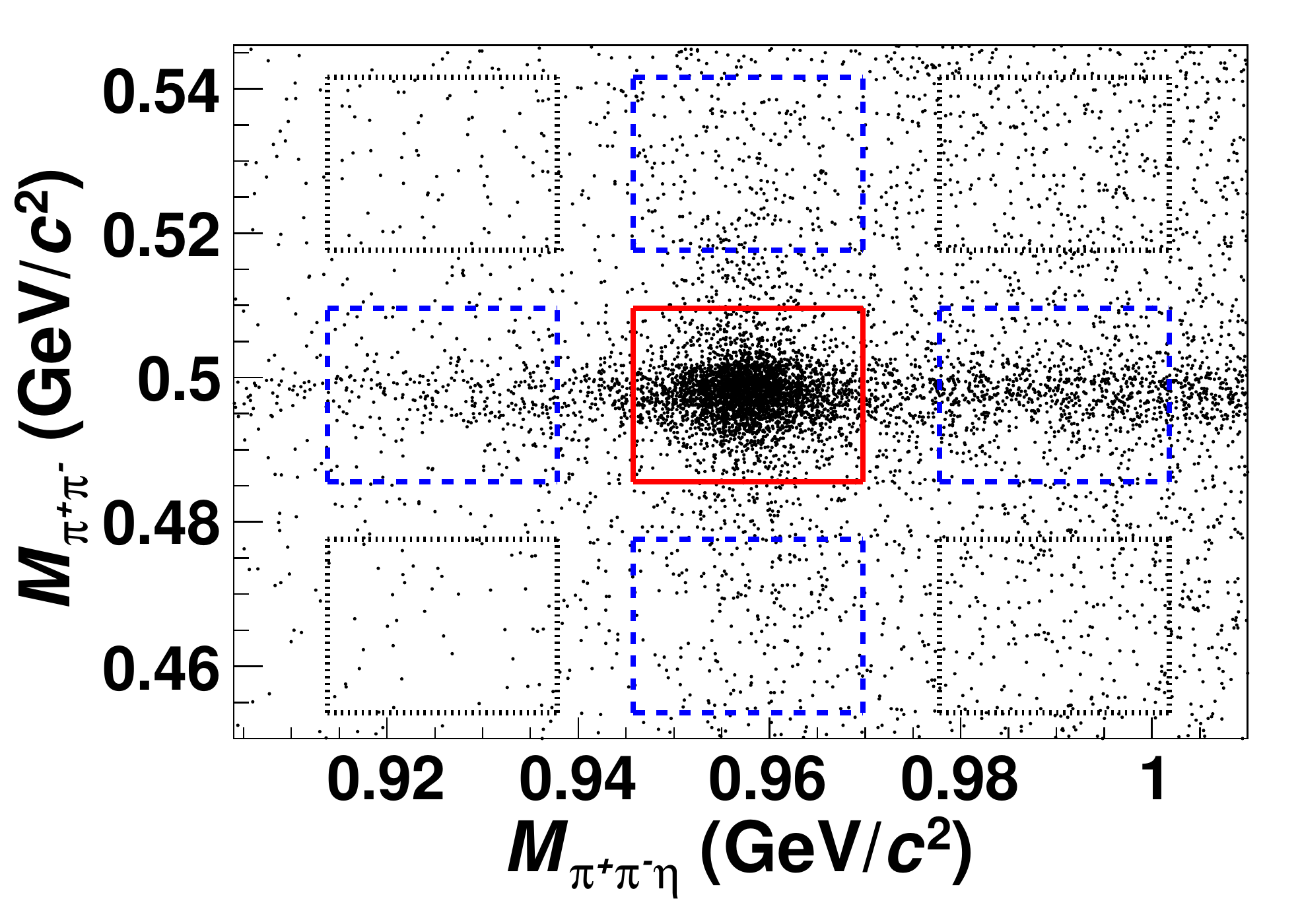}
    \end{overpic}
    }
\caption{
Distributions of (a) $M_{\pi^+\pi^-}$, (b) $M_{\pi^+\pi^-\eta}$ and
(c) $M_{\pi^+ \pi^-}$ versus $M_{\pi^+ \pi^- \eta}$ of the $D^0\to K^0_S\eta^\prime$
candidate events in data, where the regions between the pair of the solid
(dashed) arrows denote the
$K^0_S(\eta^\prime)$ signal~(sideband) regions,
the solid, dashed and dotted boxes denote the signal,
sideband A and sideband B regions (see text), respectively.
}
\label{f_side}
\end{figure*}

For the decays including $K^0_S$ $(\eta^\prime)$ mesons in the final states, there are peaking backgrounds
from non-$K^0_S$ (non-$\eta^\prime$) events in the $K^0_S$ $(\eta^\prime)$ signal regions
around the nominal $D$ mass in the $M_\text{BC}$ distributions.
To estimate these peaking backgrounds, the events in the $K^0_S$ $(\eta^\prime)$
sideband regions, defined as $0.020 <|M_{\pi^+\pi^-(\pi^+ \pi^-\eta)}-M_{K^0_S(\eta^\prime)}|<0.044\ \mathrm{GeV}/c^2$, are used.
Figure~\ref{f_side} shows the distributions of $M_{\pi^+\pi^-}$, $M_{\pi^+\pi^-\eta}$ as well as
$M_{\pi^+ \pi^-}$ versus $M_{\pi^+ \pi^- \eta}$ for the $D^0\to K^0_S\eta^\prime$ candidate events in data.
In Fig.~\ref{f_side}(a) and (b), the regions between the pair of solid~(dashed)
arrows denote the $K^0_S$ and $\eta^\prime$ signal~(sideband) regions.
To estimate the non-$K^0_S$ and non-$\eta^\prime$ peaking backgrounds in $D^0\to K^0_S\eta^\prime$ decays,
2-dimensional (2D) signal and sideband regions, as shown in Fig.~\ref{f_side}(c), are used.
The solid box is the 2D signal region, where both of the $\pi^+\pi^-$
and $\pi^+ \pi^-\eta$ combinations lie in the $K^0_S$ and $\eta^\prime$ signal regions, respectively.
The dashed~(dotted) boxes indicate the 2D sideband A~(B) regions,
in which one~(both) of the $\pi^+\pi^-$ and $\pi^+ \pi^-\eta$ combinations
lie in the $K^0_S(\eta^\prime)$ sideband regions.

The yields of peaking backgrounds in the $K^0_S(\eta^\prime)$ sideband regions in data
are obtained with similar fits to the corresponding $M_\mathrm{BC}$ distributions.
For the decays with a $K_S^0$ or $\eta^\prime$ alone in the final status, the net signal yields $N_\text{net}$ are obtained according to
\begin{equation}
    N_\text{net} = N_\text{sig} - \frac{1}{2} N_\text{sb} 
\end{equation}
where $N_\text{sig}$ and $N_\text{sb}$ are the observed number of events 
in the signal and sideband regions, respectively, 
as obtained in the fit. 
In the decay $D^0\to K_S^0\eta^\prime$, the net signal yield is estimated by
\begin{equation}
    N_\text{net} = N_\text{sig} - \frac{1}{2} N_\text{sbA} + \frac{1}{4} N_\text{sbB},
\end{equation}
where $N_{\rm sbA}$ and $N_{\rm sbB}$ denote 
the peaking background yield in the sideband regions A and B, respectively.

\section{Branching fraction}

\begin{table*}[htbp]
    \centering
    \caption{Background-subtracted signal yields ($N_\text{\bf net}$) of
    $D\to P_1P_2$ decays, the efficiencies
    ($\varepsilon$), the branching fractions measured in this work ($\mathcal B$) and
    the world average values ($\mathcal B_\text{\bf PDG}$).
    For $D^0\to P_1P_2$ decays, we include the correction factors of quantum coherence in
    $N_\text{\bf{net}}$.
    The efficiency $\varepsilon$ does not include the branching fractions
    of $\pi^0$, $\eta$, $K^0_S$ and $\eta^\prime$ decays.
    }
    \begin{ruledtabular}
    \begin{tabular}{l@{\hspace{0.5cm}}
    D{,}{\,\pm\,}{-1}@{\hspace{0.5cm}}
    D{,}{\,\pm\,}{-1}@{\hspace{0.6cm}}
    D{,}{\,\pm\,}{-1}@{$\,\pm \,$\hspace{-1.06cm}}l
    D{,}{\,\pm\,}{-1}
    }
        \multicolumn{1}{c}{Mode} & 
        \multicolumn{1}{c}{$N_\text{\bf net}$} &
        \multicolumn{1}{c}{$\varepsilon$ (\%)} & 
        \multicolumn{2}{c}{$\mathcal B$ ($\times 10^{-3}$)} & 
        \multicolumn{1}{c}{$\mathcal B_\text{\bf PDG}$ ($\times 10^{-3}$)} \\
        \colrule
        $D^+ \to \pi^+ \pi^0$                    & 10\ 108, 267 & 49.0 , 0.3 & 1.259 , 0.033 & 0.023 & 1.24 , 0.06 \\
        $D^+ \to K^+ \pi^0$                      &  1\ 822, 165 & 48.2 , 0.4 & 0.231 , 0.021 & 0.006 & 0.189, 0.025\\
        $D^+ \to \pi^+ \eta$                     & 11\ 636, 215 & 47.0 , 0.3 & 3.790 , 0.070 & 0.068 & 3.66 , 0.22 \\
        $D^+ \to K^+ \eta$                       &     439,  72 & 44.6 , 0.3 & 0.151 , 0.025 & 0.014 & 0.112, 0.018\\
        $D^+ \to \pi^+\eta^\prime$               &  3\ 088,  83 & 21.5 , 0.2 & 5.12  , 0.14  & 0.21  & 4.84 , 0.31 \\
        $D^+ \to K^+ \eta^\prime $               &      87,  25 & 18.8 , 0.2 & 0.164 , 0.051 & 0.024 & 0.183, 0.023\\
        $D^+ \to K_S^0 \pi^+$                    & 93\ 883, 352 & 51.4 , 0.2 & 15.91 , 0.06  & 0.30  & 15.3 , 0.6  \\
        $D^+ \to K_S^0 K^+$                      & 17\ 704, 151 & 48.5 , 0.1 & 3.183 , 0.029 & 0.060 & 2.95 , 0.15 \\
        \colrule
        $D^0\,\to \pi^+ \pi^-$                   & 21\ 107, 249 & 66.0 , 0.3 & 1.508 , 0.018 & 0.022 & 1.421, 0.025\\
        $D^0\,\to K^+ K^-$                       & 56\ 359, 272 & 62.8 , 0.3 & 4.233 , 0.021 & 0.064 & 4.01 , 0.07 \\
        $D^0\,\to K^\mp \pi^\pm$                 &534\ 135, 759 & 64.7 , 0.1 & 38.98 , 0.06  & 0.51  & 39.4 , 0.4  \\
        $D^0\,\to K_S^0 \pi^0$                   & 66\ 552, 302 & 37.1 , 0.2 & 12.39 , 0.06  & 0.27  & 12.0 , 0.4  \\
        $D^0\,\to K_S^0  \eta$                   &  9\ 485, 126 & 32.0 , 0.1 & 5.13  , 0.07  & 0.12  & 4.85 , 0.30 \\
        $D^0\,\to K_S^0\eta^\prime$              &  2\ 978,  61 & 12.7 , 0.1 & 9.49  , 0.20  & 0.36  & 9.5  , 0.5  \\
    \end{tabular}
    \end{ruledtabular}
    \label{tab:yields}
\end{table*}

The branching fraction of the $D\to P_1P_2$ decay is determined according to
 \begin{equation}
 {\mathcal B}(D\to P_1P_2) = \frac{N_{\rm net}}
 {2\times N^{\rm tot}_{D\bar{D}}\times \varepsilon \times{\mathcal B}},
 \label{eq:br}
 \end{equation}
where $N_{\rm net}$ is the background-subtracted signal yields of the data;
$N^{\rm tot}_{D\bar{D}}$ is the total number of 
$D\bar{D}$ pairs, which is
$(8,296 \pm 31 \pm 64) \times 10^3$ for $D^+D^-$
and $(10,597 \pm 28 \pm 87 )\times 10^3$ for $D^0\bar D^0$~\cite{hajime:ddcrs};
$\varepsilon$ is the detection efficiency obtained by the MC simulation,
and $\mathcal B$ denotes the product branching fractions of the intermediate resonances
$\pi^0$, $\eta$, $K^0_S$ and $\eta^\prime$ in the cascade decays.

The detection efficiency $\varepsilon$ is determined by analyzing the inclusive MC sample
with the same analysis
procedure as applied to the data, including the $M_\mathrm{BC}$ fit and the
background estimation. 
Because of the relatively high backgrounds in the DCS decays of $D^+\to K^+\pi^0$, $K^+\eta$ and $K^+\eta^\prime$, 
their detection efficiencies are determined from MC samples of $\psi(3770)\to D^+D^-$ in which 
one $D$ is forced to decay into a signal mode and the other decays generically. 
By fitting the $M_\text{BC}$ distributions we obtain the net signal
yield from the MC samples for each decay. 
The detection efficiency is obtained by dividing the net signal yield by the total number of the produced 
signal events. 
To better describe the data, the MC simulated efficiencies are corrected by the differences between data and MC simulation as discussed in
Sect.~\ref{sec_sys}.

Inserting the values of $N_{\rm net}$, $N^{\rm tot}_{D\bar{D}}$,
$\varepsilon$ and $\mathcal B_i$ in Eq. (\ref{eq:br}), we obtain the
branching fractions of the decays of interest, as listed in Table~\ref{tab:yields}.
For the branching fractions measured in this work, the first uncertainty is statistical
and the second one is systematic.
By subtracting the branching fraction of  
DCS decay $D^0 \to K^+\pi^-$~\cite{pdg2016review}
from that of $D^0 \to K^\mp\pi^\pm$, 
we obtain the branching fraction of $D^0 \to K^-\pi^+$ to be $(3.882 \pm 0.006 \pm 0.051)\%$.

\section{Systematic uncertainty} \label{sec_sys}

Table~\ref{tab:sys} summarizes the sources of the systematic uncertainties
in the branching fraction measurements. The uncertainties are estimated relative to
the measured branching fractions and are described below.

\begin{itemize}
\item
$\bm{N^{\rm tot}_{D\bar{D}}}$:
The total number of $D\bar{D}$ pairs produced in data 
are cited from our previous
work~\cite{hajime:ddcrs}. They are determined with a combined analysis
in which both single-tag and double-tag events are used.
Their uncertainties are included in our measurement.

\item
{\bf Tracking (PID) of $K^+(\pi^+)$}:
The tracking (PID) efficiencies for $K^+(\pi^+)$
are studied by using double-tag $D\bar D$ hadronic events.
Small differences in the tracking (PID) efficiencies of $K^+(\pi^+)$
between data and MC simulation (denoted as data-MC differences) have been observed.
To better describe the data, the MC simulated efficiencies are corrected by the momentum dependent data-MC differences for the $K^+$ or ${\pi^+}$.
Afterwards, the systematic uncertainty for tracking (PID) is assigned as
1.0\% (0.6\%) for each pion from $\eta^\prime$ decays, and
0.3\% (0.3\%) per track for the others.

\begin{table*}[htbp]
\caption{Relative systematic uncertainties (in \%) in the branching fraction measurements.}
\centering
\begin{ruledtabular}
\begin{tabular}{lcccccccccccccc}
Source & $\pi^+ \pi^0$ & $K^+ \pi^0$ & $\pi^+ \eta$ & $K^+ \eta$ & $\pi^+ \eta^\prime$
& $K^+ \eta^\prime$& $K_S^0 \pi^+$& $K_S^0 K^+$
& $\pi^+ \pi^-$ &  $K^+ K^-$ &$K^\mp \pi^\pm$ & $K_S^0 \pi^0$ &  $K_S^0
\eta$ &$K_S^0 \eta^\prime$ \\ \hline

$N^{\rm tot}_{D\bar D}$       &0.9&0.9&0.9&0.9&0.9& 0.9&0.9&0.9&0.9&0.9&0.9&0.9&0.9&0.9\\
Tracking of $K^+(\pi^+)$      &0.3&0.3&0.3&0.3&2.3& 2.3&0.3&0.3&0.6&0.6&0.6& - & - &2.0\\
PID of $K^+(\pi^+)$           &0.3&0.3&0.3&0.3&1.5& 1.5&0.3&0.3&0.6&0.6&0.6& - & - &1.2\\
$K_S^0$ reconstruction        & - & - & - & - & - &  - &1.5&1.5& - & - & - &1.5&1.5&1.5\\
$\pi^0(\eta)$ reconstruction  &1.0&1.0&1.0&1.0&1.0& 1.0& - & - & - & - & - &1.0&1.0&1.0\\
$\Delta E$ requirement        &0.1&1.2&0.6&2.6&0.5& 3.8&0.3&0.2&0.5&0.6&0.2&0.4&0.4&0.4\\
$M_\text{BC}$ fit             &0.9&1.7&0.5&8.5&1.8&13.3&0.2&0.2&0.6&0.4&0.1&0.2&0.4&0.4\\
Background estimation         & - & - & - & - &0.2& 4.3&0.1&0.3& - & - & - &0.5&0.2&0.8\\
Quoted\,branching\,fractions  &0.0&0.0&0.5&0.5&1.7& 1.7&0.1&0.1&  -&  -&  -&0.1&0.5&1.7\\
MC statistics                 &0.7&0.8&0.5&0.6&0.8& 1.0&0.3&0.3&0.4&0.5&0.1&0.5&0.4&0.6\\
QC effects                    & - & - & - & - & - &  - & - & - &0.2&0.1&0.1&0.2&0.5&0.7\\ 
\hline  
Total                         &1.8&2.6&1.8&9.0&4.1&14.9&1.9&1.9&1.5&1.5&1.3&2.2&2.3&3.8\\
\end{tabular}
\end{ruledtabular}
\label{tab:sys}
\end{table*}

\item
{\bf $K^0_S$ reconstruction}:
The $K_{S}^{0}$ reconstruction efficiency,
including the tracking efficiency for charged pions,
is studied with control samples of
$J/\psi\to K^{*}(892)^{\mp}K^{\pm}$ 
with $K^{*}(892)^{\pm} \to K_S^0 \pi^\pm$
and $J/\psi\to \phi K_S^{0}K^{\pm}\pi^{\mp}$.
Small data-MC differences are found, as presented in Ref.~\cite{ablikim2015study}.
We correct the MC efficiencies for these differences and assign a systematic uncertainty of $1.5\%$ 
for each $K^0_S$.

\item
{\bf $\pi^0$ and $\eta$ reconstruction}:
The $\pi^0$ reconstruction efficiency is verified with double-tag
hadronic events
$D^0\to K^-\pi^+$ and $K^-\pi^+\pi^+\pi^-$ versus $\bar{D^0}\to K^-\pi^+\pi^0$
and $K_{S}^{0}(\pi^+\pi^-)\pi^0$.
Small data-MC differences for the $\pi^0$ reconstruction efficiencies
are found and 
are corrected to the MC simulation efficiencies.
After corrections, the uncertainty for the $\pi^0$ reconstruction efficiency is
taken as 1.0\%.
The uncertainty for the $\eta$ reconstruction efficiency is assigned as 1.0\%, too.

\item
{\bf $\Delta E$ requirement and $M_{\rm BC}$ fit}:
The uncertainty from the $\Delta E$ requirement is investigated with alternative
requirements of $3.5\,\sigma_{\Delta E}$ or $4.0\,\sigma_{\Delta E}$. The resultant largest changes
in the branching fractions are assigned as the uncertainties.
The uncertainty from the $M_{\rm BC}$ fit is examined 
with different fit ranges $(1.8335, 1.8865)$ or
$(1.8395, 1.8865)\ \mathrm{GeV}/c^2$, different endpoints
of $1.8863$ or $1.8867\ \mathrm{GeV/}c^2$ for
the ARGUS function, and different signal shapes with various requirements on
the MC-truth matched signal shapes.  
The largest changes on the branching fractions with respect to the nominal results
are treated as the corresponding systematic uncertainties.

\item
{\bf Background estimation}:
The uncertainty from the $K_S^0(\eta^\prime)$
sideband region is examined by changing the scale factors based on MC simulations
and by shifting the $K^0_S(\eta^\prime)$ signal or sideband
regions by $\pm 2\ \mathrm{MeV}/c^2$.
The maximum changes of
the branching fractions
with respect to the nominal results
are assigned as the systematic uncertainties due to
background estimation.

For the decay of $D^+\to K^+ \pi^0$, we also examine the effect of the fixed
peaking background of $D^+\to K^0_S(\to\pi^0\pi^0)\pi^+$
by considering the uncertainties of 
its world average branching fraction~\cite{pdg2016review},
the tracking and PID for $\pi^+$ and the $\pi^0$ selection. The effect is found to be negligible.

\item
{\bf Quoted branching fractions}:
The uncertainties in the quoted branching fractions for
$\pi^0\to \gamma\gamma$, $\eta \to \gamma\gamma$, $K_S^0 \to \pi^+ \pi^-$ and
$\eta^\prime \to \pi^+\pi^-\eta$ are $0.03\%$, $0.51\%$, $0.07\%$ and
$1.63\%$~\cite{pdg2016review}, respectively.

\item
{\bf MC statistics}:
The uncertainty in the efficiencies due to limited MC statistics is
taken into account.

\item
{\bf Quantum coherence (QC) effects}:
Since $D^0$ and $\bar D^0$ are coherently produced in the process $e^+e^-\to\psi(3770)\to D^0\bar D^0$, 
quantum correlation is considered with 
a method introduced in Ref.~\cite{asner2006time} when measuring the signal yields.
The correction factors are included in the signal yields listed in Table~\ref{tab:yields}.
The parameters are quoted from the PDG~\cite{pdg2016review} and Heavy Flavor Averaging Group~\cite{www:www.slac.stanford.edu}
and their uncertainties propagate to the branching fractions as systematic uncertainties.

\end{itemize}

Assuming all the uncertainty sources are independent, the 
quadratic sum of these uncertainties gives
the total systematic
uncertainty in the measurement of the branching fraction for each decay.

\section{Summary}

By analyzing the data sample corresponding to an integrated luminosity of 2.93 fb$^{-1}$ taken at $\sqrt s= 3.773\ \mathrm{GeV}$,
we measure the absolute branching fractions for the two-body hadronic decays
$D^+\to \pi^+ \pi^0$,
$K^+ \pi^0$, $\pi^+ \eta$, $K^+ \eta$, $\pi^+ \eta^\prime$, $K^+ \eta^\prime$,
$K_S^0 \pi^+ $, $K_S^0 K^+$, and $D^0\to \pi^+ \pi^-$,
$K^+ K^-$, $K^- \pi^+$, $K_S^0 \pi^0$, $K_S^0 \eta$, $K_S^0 \eta^\prime$.
As shown in Table~\ref{tab:yields}, our results are consistent with the
world average values within uncertainties and
the branching fractions of $D^+\to\pi^+\pi^0$, $K^+\pi^0$, $\pi^+ \eta$,  $\pi^+ \eta^\prime$,
$K_S^0 \pi^+$, $K_S^0 K^+$ and $D^0 \to K_S^0 \pi^0$, $K_S^0 \eta$, $K_S^0 \eta^\prime$ are determined with improved
precision.
The measured branching fractions for $D^{0}\to K^0_S\pi^0$ and $D^+\to K^0_SK^+$ 
are consistent with those measured using a double-tag technique in our previous works~\cite{jingzhen:ks},
but with better precision.
These results are useful for tests of theoretical calculations
and provide a better understanding of SU(3)-flavor symmetry breaking effects in 
hadronic decays of the $D$ mesons~\cite{waikwok1993minimal, Grossman2013,yu2011nonleptonic,li2012branching}.

\section{Acknowledgements}
\input{acknowledgement}

\end{document}

%% file: author_final.tex
\author{
    \small
    M.~Ablikim$^{1}$, M.~N.~Achasov$^{9,d}$, S. ~Ahmed$^{14}$, M.~Albrecht$^{4}$, A.~Amoroso$^{53A,53C}$, F.~F.~An$^{1}$, Q.~An$^{50,40}$, J.~Z.~Bai$^{1}$, Y.~Bai$^{39}$, O.~Bakina$^{24}$
% ...
    \\
    \vspace{0.2cm}
    (BESIII Collaboration)\\
    \vspace{0.2cm} {\it
    $^{1}$ Institute of High Energy Physics, Beijing 100049, People's Republic of China\\
% ...
$^{4}$ Bochum Ruhr-University, D-44780 Bochum, Germany\\
% ...
$^{9}$ G.I. Budker Institute of Nuclear Physics SB RAS (BINP), Novosibirsk 630090, Russia\\
% ...
$^{14}$ Helmholtz Institute Mainz, Johann-Joachim-Becher-Weg 45, D-55099 Mainz, Germany\\
% ...
$^{24}$ Joint Institute for Nuclear Research, 141980 Dubna, Moscow region, Russia\\
% ...
$^{39}$ Southeast University, Nanjing 211100, People's Republic of China\\
    $^{40}$ State Key Laboratory of Particle Detection and Electronics, Beijing 100049, Hefei 230026, People's Republic of China\\
% ...
$^{50}$ University of Science and Technology of China, Hefei 230026, People's Republic of China\\
% ...
$^{53}$ (A)University of Turin, I-10125, Turin, Italy; (B)University of Eastern Piedmont, I-15121, Alessandria, Italy; (C)INFN, I-10125, Turin, Italy\\
% ...
$^{d}$ Also at the Novosibirsk State University, Novosibirsk, 630090, Russia\\ 
% ...
}
    \vspace{0.4cm}
}

%% file: acknowledgement.tex
The BESIII collaboration thanks the staff of BEPCII and the IHEP computing center for their strong support. 
This work is supported in part by National Key Basic Research Program of China under Contract No. 2015CB856700; 
National Natural Science Foundation of China (NSFC) under Contracts 
Nos. 11475090, 11305180, 10975093, 11005061, 11235011, 11335008, 11425524, 11475107, 11625523, 11635010; 
the Chinese Academy of Sciences (CAS) Large-Scale 
Scientific Facility Program; the CAS Center for Excellence in Particle Physics (CCEPP); 
Joint Large-Scale Scientific Facility Funds of the NSFC and CAS under Contracts Nos. U1632109, U1332201, U1532257, U1532258; 
CAS under Contracts Nos. KJCX2-YW-N29, KJCX2-YW-N45; 
CAS Key Research Program of Frontier Sciences under Contract No. QYZDJ-SSW-SLH003;
100 Talents Program of CAS; National 1000 Talents Program of China; INPAC and Shanghai Key Laboratory for Particle Physics and Cosmology; German Research Foundation DFG under Contracts Nos. Collaborative Research Center CRC 1044, FOR 2359; Istituto Nazionale di Fisica Nucleare, Italy; 
Joint Large-Scale Scientific Facility Funds of the NSFC and CAS; 
Koninklijke Nederlandse Akademie van Wetenschappen (KNAW) under Contract No. 530-4CDP03; Ministry of Development of Turkey under Contract No. DPT2006K-120470; 
National Natural Science Foundation of China (NSFC); 
National Science and Technology fund; 
The Swedish Resarch Council; U. S. Department of Energy under Contracts Nos. DE-FG02-05ER41374, DE-SC-0010118, DE-SC-0010504, DE-SC-0012069; University of Groningen (RuG) and the Helmholtzzentrum fuer Schwerionenforschung GmbH (GSI), Darmstadt; WCU Program of National Research Foundation of Korea under Contract No. R32-2008-000-10155-0.